\documentclass[aps,twocolumn]{revtex4}
\usepackage{graphics}
\usepackage{psfrag}
\usepackage{times}
\usepackage{amsmath, amssymb}
\newcommand{\pd}{\phantom{\dagger}}
\begin{document}
\title{ A simple model to explain narrow nucleon resonances below the
                    $\pi$ threshold}
\author{ 
          Thomas~Walcher
  }
\affiliation{
  Institut f\"ur Kernphysik, Johannes Gutenberg-Universit\"at Mainz,
  D-55099 Mainz, Germany 
  }
\date{\today}
\begin{abstract}
Several experiments suggest the existence of narrow excited states in
the nucleon below the $\pi N$ threshold. Taking all information
together a series of states emerges at masses of 940, 966, 985, 1003,
$m_4$, 1044, $m_6$, $1094\,MeV$ above the nucleon ground state at
$940\,MeV$ with $m_4$ and $m_6$ missing. A model based on the
excitation of collective states of the quark condensate is proposed
explaining these states as multiple production of a ``genuine''
Goldstone Boson with  a mass of $\approx 20\,MeV$. The model also
accounts in a natural way for the non-observation of $m_4$ and $m_6$. 
It suggests an explanation of the mass gap problem of the hadronic 
spectrum and a confinement mechanism.  
\end{abstract}


\maketitle
\section{Introduction}                                            \label{intro}
Recently two experiments reported about narrow exited states of the
neutron below the $\pi$ threshold \cite{Tat97,Fil00,Bec01a}. In the
first \cite{Tat97} the reaction $pp \rightarrow p \pi^+ N^*$ has been   
investigated at three energies of  $T_p = 1520, 1805$, and
$2100\,MeV$. Three peaks about $5\,MeV$ wide at $1004, 1044$, and 
$1094\,MeV$ were observed in the covered mass range 
$960 < m_{N^*} < 1170\,MeV$ with a statistical significance between 
17 and 2 standard deviations. In the second experiment \cite{Fil00} 
also three peaks are found in a partially overlapping mass range 
in the $pd \rightarrow p p X_1$ reaction with 
$X_1 = N^* \rightarrow N \gamma$ at a $p$ energy 
of $305\,MeV$. From a measurement of the $pp$ the missing mass of 
$X1$ at $966\pm2, 985\pm2$ and $1003\pm2\,MeV$ were determined with 
an experimental width of $5\,MeV$ and about 6 standard 
deviations statistical significance. These peaks were assigned to
``super narrow dibaryons'' states, but can as well be interpreted as 
narrow excited states $X1$ of the nucleon. A third evidence comes 
from a measurement of the $\gamma p\rightarrow p\gamma$ Compton 
scattering in the photon-energy range $60<E_{\gamma}<160\,MeV$ where a  
peak with an experimental resolution of $5~MeV$ at $\approx 1048\,MeV$ 
with 3.5 standard deviations is observed \cite{Bec01b}. In the
charge exchange reaction $pp \rightarrow n X^{++}$, however, no
doubly charged excited nucleons $X^{++}$ could be found \cite{Ram94}. 
Since such narrow states in the nucleon are very surprising indeed, 
they meet a considerable scepticism. 

In this paper these peaks are taken serious as narrow
states of the nucleon and a model for their explanation is proposed. 
The model starts from the observation that the series of 
masses taking all experiments together and including the neutron
ground state gives $940, 966, 985, 1003, m_4, 1044, m_6, 1094\,MeV$
with $m_4$ and $m_6$ missing. If one completes the series with 
$m_4=1023\,MeV$ and $m_6=1069\,MeV$, then the masses are equidistant
within the errors with an average mass difference of 
$\delta m=21.2\pm2.6\,MeV$.
This mass difference is close to the best guesses of
two times the mass of the light current quarks $m_q=8 \cdots 14\,MeV$ 
\cite{Don92,Tho01}. Somewhat against the current thinking we
hypothesize the existence of a light pseudo scalar meson 
($J^P = 0^-$), i.e. a light ``$\pi$'', with a mass of $m_{light\,\pi}= 
2 m_{q} = 21\,MeV$. The basic idea is now that the series of excited 
states is due to the nucleon in its ground state plus  $1,2,3, \dots$
light $\pi$s as the quantum of excitation with the energy
$m_{light\,\pi}$.  

The pseudo scalar character of the light $\pi$ explains 
naturally the missing of the states \#4 and \#6. The experiment in 
ref.~\cite{Tat97} detected the $p$ and $\pi^+$ in the 
$pp \rightarrow p \pi^+ N^*$ reaction in one spectrometer with a very
small relative angle at the same time. Since the cross section peaks 
very strongly in forward direction with respect to the beam 
($\theta \approx 5^0$),  this means that the outgoing $p$ and 
$\pi$ carry no angular momentum and only odd parity states in the 
$N^*$ can be excited. On the other hand, in the 
$pd \rightarrow ppX_1$ reaction the proton on the left hand side was 
measured at $\theta_L=70^0$ and the proton on the right hand side at 
angles of $\theta_R=34^0, 36^0$, and $38^0$. Therefore, odd and even
partial waves of the outgoing particles are possible and allow the
population of odd and even parity states in $N^*$. The presence of the
second nucleon in the $pd \rightarrow ppX_1$ reaction may provide
the effect of the $\pi$ present in the first reaction. The peak
at $m_{N^*}\approx 1048\,MeV$ in the $\gamma p \rightarrow p \gamma$
reaction is a $J^P = 1/2^-$ state and due to an $E1$ transition 
dominating at low energies $E_{\gamma}$. All excited nucleons 
suggested in these experiments have positive or no charge $N^{*0,+}$
and can decay electromagnetically. On the other hand, the
$N^{*++}=X^{++}$ in the experiment of ref.~\cite{Ram94} could decay 
only weakly and had a survival time $\tau_{N^{*++}}\approx 10^{-3}s$.
This feature was used to search for the $N^{*++}$ directly in a
magnetic spectrometer, but as already mentioned with negative result. 
As will be discussed in section~\ref{discussion} the light $\pi$
couples only very weakly and has a small overlap with the usual $\pi$
making it unlikely to excite the $N^*$ states in the $pp$ charge
exchange reaction.  

However, it is well known that a light $\pi$ has never been 
observed as free particle and one has to give more
justification for this proposal and reconcile it with the known
physics. One fact of this known physics is the success of the
constituent quark model (see e.g. ref.~\cite{Tho01} and references
therein). This model uses ``constituent'' quarks with the naive
estimate of the quark mass $m_c \approx 350\,MeV$, i.e. one third of 
the nucleon mass. Therefore, the $\pi$ had to have as any other meson 
at least a mass of $\gtrapprox 700\,MeV$. However, the $\pi$ observed 
in nature has a mass of $138\,MeV$. The model described in the 
following assumes that the $\pi$ observed in nature is not the 
Goldstone Boson, it is normally identified with, but a mixture of the 
``genuine'' Goldstone Boson $\pi_G$, i.e. the light $\pi$ suggested  
above consisting of light current quarks, with a ``constituent'' 
$\pi_c$ made of constituent quarks with a mass of 
$m_{\pi_c} = 2 m_{c}= 700\,MeV$. It will turn out that the mixture of 
these states will reproduce the observed $\pi$ mass and explain in a 
natural way the so called ``gap problem'', i.e. the large 
mass gap between the mass of the hadronic scale of 
$m_{\rho} \approx 2 m_{c}$ and the mass of the observed $\pi$ 
(see e.g. \cite{Tho01}). 

\section{Model}                                             \label{Model}
\begin{figure}[ht]
  \begin{center}
   \psfrag{i}[t][][3.]{$i$}
   \psfrag{j}[b][][3.]{$j$}  
   \psfrag{m}[t][][3.]{$m$}
   \psfrag{n}[b][][3.]{$n$}  
  \resizebox{0.9\columnwidth}{!}
               {\includegraphics{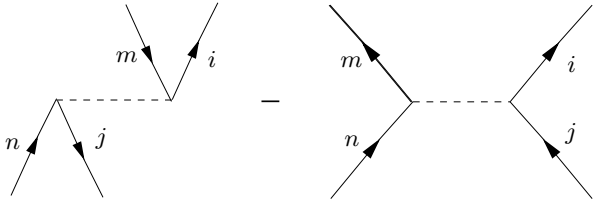}}
  \end{center}
  \caption{Graphical representation of particle-hole states.}
  \label{fig:diagrams}
\end{figure}

The model follows the formulation of the schematic model of
G.E.~Brown \cite{Bro} for the description of collective states in
nuclei as particle-hole excitation and this paper adopts closely his  
notation. The Goldstone Boson is denoted by
$\pi_G = |q_{curr.}\bar{q}_{curr.}\rangle = |mi\rangle$ and the
constituent pion by
$\pi_c = |q_{const.}\bar{q}_{const.}\rangle = |nj\rangle$. They
represent particle-hole excitations of the QCD vacuum where $m,n$
label the particle and $i,j$ the hole states. The possible
propagation of the particle-hole states are depicted in
fig.\ref{fig:diagrams}. 
The model assumes the suppression of the closed Fermion loop coupling
to the particle-hole state inside the confinement cavity. This
assumption will be discussed in section~\ref{discussion}, item 1) . 
The particle-hole interaction is described by a non relativistic
Hamiltonian  
\begin{multline}
H = \sum_{k_1,k_2}\langle k_1|T|k_2 \rangle 
               a^{\dagger}_{k_1}a^{\pd}_{k_2} +\\+ 
    \frac{1}{2}\sum_{k_1,k_2,k_3,k_4}\langle k_1k_2|V|k_3k_4 \rangle 
       a^{\dagger}_{k_1}a^{\dagger}_{k_2}a^{\pd}_{k_1}a^{\pd}_{k_2}
\label{eq:H} 
\end{multline} 
where the $a$ and $a^{\dagger}$ are the creation and annihilation
operators, respectively. The kinetic energy operator $T$ is a one body
and the particle-hole interaction $V$ a two body operator. The binding
energies of the quarks to $\pi$s is much smaller than their mass as will
be discussed below in section~\ref{discussion}, item 3), justifying
the non relativistic treatment.The off-diagonal matrix elements of the
particle-hole interaction is then given by
\begin{equation}
\langle mi|H|nj \rangle = \langle mj|V|in\rangle - \langle jm|V|in\rangle.
\end{equation}
For the diagonal matrix elements one gets
\begin{equation}
\langle mi|H|mi \rangle = (\epsilon_m - \epsilon_i) + 
                      \langle mi|V|im\rangle - \langle im|V|im\rangle
\end{equation}
The particle-hole single particle energies are here just given by the
total energy of the $q\bar{q}$ pair created out off the vacuum, i.e.
\begin{equation}
m_{mi} = \epsilon_m - \epsilon_i
\end{equation}
The particle-hole interaction mixes the particle-hole states as
already mentioned. In order to find the energy eigenvalues $E$ of the
mixed states one has to solve the secular equation:
\begin{equation}
\sum_{n,j}\langle mi|H|nj \rangle c_{nj} = E c_{mi}
\end{equation}

For the determination of $E$ one observes that only the left hand
diagram in fig.~\ref{fig:diagrams} contributes essentially. The
right hand diagram is an exchange diagram propagating the energy 
equivalent to the mass difference of the constituent quark to the 
current quark. Since this difference is of the order $700\,MeV$ the
exchange matrix element will be small compared to the particle-hole
creation and annihilation matrix element which does not propagate
energy or momentum. It is now assumed that the $S$ matrix represented
by the particle-hole propagation is unitary. Consequently, the
creation-annihilation matrix element factorizes according to:
\begin{equation}
\langle mj|V|in\rangle = 
   \lambda \langle mj|\hat{V}|0\rangle \langle 0|\hat{V}|in\rangle
\end{equation}
where $\hat{V}$ represents the $q\bar{q}$ transition at the vertices
in fig.~\ref{fig:diagrams}. 
This factorization is familiar from the more complete RPA description  
of collective particle-hole excitations in nuclei.
The factor $\lambda$ takes care of the slight violation of unitarity
due to the neglect of the exchange diagram. The absolute squares of 
the matrix elements 
$|\langle mj|\hat{V}|0\rangle|^2 = m_{mj}$ and 
$|\langle 0|\hat{V}|in\rangle|^2 = m_{ni}$ are just the masses
of the particle-hole states. 

\begin{figure}[ht]
  \begin{center}
   \psfrag{p1}[t][][1.]{$E$}
   \psfrag{p2}[b][][1.]{} 
  \resizebox{0.9\columnwidth}{!}
               {\includegraphics{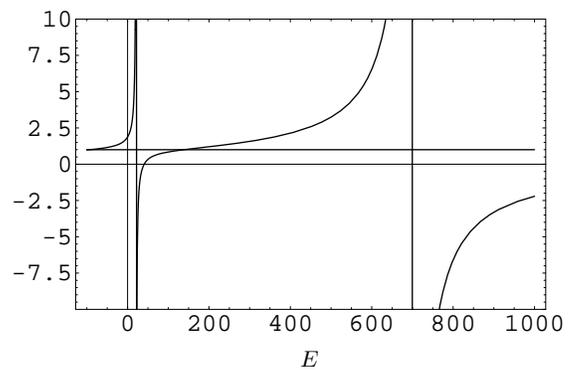}}
  \end{center}
  \caption{Graphical solution of the secular equation (\ref{eq:sec}).}
  \label{fig:sec}
\end{figure}

Putting all equations together one arrives at the secular equation:
\begin{equation}
1 = \sum_{m,i}\frac{\lambda m_{mi}}{E-(\epsilon_m - \epsilon_i)} = 
    \sum_{m,i}\frac{\lambda m_{mi}}{E-m_{ij}} 
\end{equation}
For the two possible particle-hole states mixing here the secular
equation takes finally the simple form:
\begin{equation}
1=\frac{\lambda m_G}{E-m_G} + \frac{\lambda m_c}{E-m_c} 
\label{eq:sec}  
\end{equation}
The solution of this equation is most easily visualized by a plot of
the right hand side and the left hand side as a function of $E$. This
plot is shown in fig.~\ref{fig:sec}.  
The physical solution appears at $E=m_{\pi}=138\,MeV$ with the fitted 
$\lambda=-0.94$ if one chooses $m_G=21\,MeV$ and $m_c=700\,MeV$ as 
discussed above. Since $\lambda \approx -1$ this solution shows that 
the creation-annihilation diagram dominates indeed and that the
particle-hole interaction is attractive as required.

\section{Discussion}                                     \label{discussion}
It is evident that the proposed identification of the Goldstone Boson
having $J^P = 0^-$ explains the combined experimental results for the
narrow nucleon states presented above. If the available energy is
below the $\pi$-nucleon threshold one can just see the
excitation of a series of ``genuine''
Goldstone Bosons in the nucleon spectrum. In other words the energy
available below the $\pi$ threshold and the binding of $\pi_G$ to the
nucleon cause a de-mixing of the Goldstone $\pi_G$ and the constituent
$\pi_c$. Narrow mass steps could of course not be seen in the
meson spectra since already the ground state of the mesons decay in 
normal $\pi$s and, therefore, the Goldstone $\pi_G$ mixes with the 
constituent $\pi_c$ through the energy available in the decay.

\begin{figure}[ht]
  \begin{center}
   \psfrag{p1}[t][][1.]{$n$}
   \psfrag{p2}[b][][1.]{dibaryon mass [$MeV$]}  
  \resizebox{0.9\columnwidth}{!}
               {\includegraphics{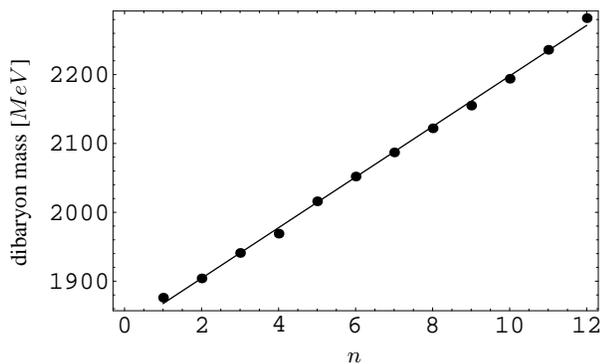}}
  \end{center}
  \caption{The experimental dibaryon masses according to
   ref.~\cite{Tat99} against the number of excitation quanta $n$.}
  \label{fig:spectrum} 
\end{figure}

It is interesting to consider also the narrow dibaryon states claimed
so far since they could be naturally explained as narrow bound states
composed of $NN^*$ or $N^*N^*$. A compilation of the references of
these states and a mass spectrum derived from them is given in
ref.~\cite{Tat99}. It is striking that this spectrum of narrow
dibaryon states shows a rather equidistant level spacing. 
Fig.~\ref{fig:spectrum} depicts this spectrum after 
ref.~\cite{Tat99} together with a straight line fit to the masses 
$m = 1867.5\,MeV + n \cdot 36.7\,MeV$ where $n$ is
the number of steps with the excitation quantum $36.7\,MeV$. The
probability that the observed spectrum is found accidentally is given 
in good approximation by $W=(\sigma/\Delta m)^{(n-1)} \approx
10^{-20}$ where  $\sigma =6\,MeV$ is the variance of the observed
masses with respect to the hypothetical equidistant spectrum and 
$\Delta m = 406\,MeV$ is the mass interval considered. It is important  
to note that the authors of ref.\cite{Tat99} had no bias to a 
hypothesis of equidistant dibaryon masses. 
The level spacing is very close to two times the mass of the assumed
light Goldstone Boson though no attempt has been made to correct for
the effect of the binding energy of the dibaryons. 
As already argued the proposed model implies that in most experiments 
only the natural parity states are excited so that the unnatural
parity states are missing and the step width is indeed two times the 
mass of the excitation quantum, i.e. the light Goldstone Boson mass. 

The model implies some noteworthy further conclusions:

1) The free $\pi$ observed in nature is not the true Goldstone Boson, 
but a mixed state of the ``genuine'' Goldstone Boson composed of  
current quarks with a mass $m_q \approx 10\,MeV$ and the constituent
$\pi_c$ composed of constituent quarks with a mass of $m_c \approx
350\,MeV$. This idea represents a modification of the usual 
dynamical generation of the $\pi$ mass through coupling of the 
particle-hole to the quark condensate $\langle 0|q\bar{q}|0 \rangle$
in the Nambu-Jona-Lasinio (NJL) model (see e.g. ref.~\cite{Vog91,Tho01}. 
The NJL model does not know confinement and a coupling
to the closed Fermion loop, i.e. to the quark condensate, is always 
possible. In order to explain the narrow $N^*$ states, however, one
had to assume that the coupling of the particle-hole state to the 
closed Fermion loop disappears inside the confinement cavity. This 
meant that the quark condensate is ``expelled'' out off the nucleon 
leaving a cavity with a real vacuum inside as suggested by the 
``color diaelectric model'' (see e.g. ref.~\cite{Tho01}).\\    
More generally, the quark condensate as the leading order parameter of
the spontaneously broken symmetry is a theoretical concept and is not
yet experimentally established \cite{Col01}. 

2) The Goldstone Boson does not interact at all in the chiral limit 
$m_q \rightarrow 0$ (see e.g. \cite{Tho01}) and it is reasonable to
assume that the ``genuine'' Goldstone Boson $\pi_G$ interacts only
very weakly. This assumption explains why the narrow states have not
been observed in less sensitive Compton scattering experiments than
that of ref.~\cite{Bec01b} as already remarked in ref.~\cite{Lvo98}.
Due to this very weak coupling $\pi_G$ will have very little influence
in dispersion relations, as e.g. for the form factors, or on
polarizabilities of the $\pi$ and nucleon.\\
The chiral loops in the framework of the Chiral Perturbation Theory
(ChPTh) (see e.g. ref.~\cite{Don92,Tho01}) have, however, to be 
reconsidered. Since the $\pi$ mass is an effective parameter in ChPTh 
and only requested to be small compared to the hadronic scale
$m_{\rho} \approx 700\,MeV$ it appears possible that a consistent
picture also with $\pi_G$ emerges. In this modification of ChPTh the 
coupling constants and low energy constants had to be changed. It 
could also be that the $\pi$ mass in ChPTh is a more delicate
parameter than realized so far. For observables with external $\pi$s  
one had to take the asymptotic mass $m_{\pi}=138~MeV$ with strong 
coupling. In the chiral loops one either has to modify the low energy 
constants for the ``genuine'' Goldstone $\pi$s with 
$m_G \approx 20\,MeV$ or had to take the effective $\pi$s with 
$m_{\pi}$ produced by a self consistent condensate of $\pi_G$.
 
3) The binding energy of a system of two current quarks $q\bar{q}$ 
by the familiar one gluon exchange potential 
$V(r) = (4/3) \alpha_s(Q^2)/r$ to $\pi_G$ is $\epsilon_B=-2\,MeV$ 
if one uses the saturation value $\alpha_s=0.5$ for 
$Q \leqslant 2\,GeV/c$ suggested by the analysis of jet shapes 
\cite{Dok01}. This value of $\alpha_s$ is low in comparison to 
the running coupling value and means that beside $\Lambda_{QCD}$ 
a second scale is introduced. Consequently, the $\pi_G$ would be a 
relatively large object with a size estimated by the Compton wave 
length $\lambdabar_{Compton}=\hbar/21\,MeV/c \approx 10\,fm$.
This means that the correlation length of chiral symmetry would be  
much larger than the best estimate of the confinement radius of 
$R_{conf.}\approx 1\,fm$ \cite{Pov86}. Therefore, the overlap 
of the $N^* = (n\cdot\pi_G) N\,$ states with the ground state of the
nucleon is small and together with the very weak $\pi_GN$ coupling
gives a further reason why it is so difficult to see the $\pi_G$
in experiments. 

4) The mechanism proposed in this paper provides for a natural 
explanation of the ``gap problem''. It is, however, unclear how this
relates to deeper considerations of this problem
\cite{Kon01a,Kon01b}. 

5) The existence of a light Goldstone Boson would also provide
a mechanism for the quark confinement. A free non-confined quark would 
be naturally a constituent quark, i.e. one dressed with gluons and a 
mass of $m_c \approx 350\,MeV + m_{q}$. The binding energy
$\epsilon_B$ of a system of $q_c\bar{q}_c$ by the mentioned one gluon 
exchange potential is  $\epsilon_B=-32\,MeV$ if one uses the saturated 
value $\alpha_s=0.5$ of ref.~\cite{Dok01}. This is certainly a lower
bound of $|\epsilon_B|$. If $|\epsilon_B|$ is larger than the mass 
$m_G$ of the ``genuine'' Goldstone Boson a quark could never escape
since before it did, as many Goldstone Bosons as energy is available
would be produced, effectively taking away the energy of the quark
above the ground state. Since the Goldstone Bosons interact only 
very weakly with the colored constituents the influence on the quark 
binding potential would be weak. Before, however, such a picture 
can be established self consistent calculations including all 
effective degrees of freedom, i.e. the constituent quarks, the 
Goldstone Bosons and the gluons had to be performed.

6) The lattice gauge theory allows for such a self consistent 
calculation, but it may miss the physics of the effective degree
of freedom represented by the Goldstone Boson because its correlation
length is, as argued, $10~fm$ and all lattices used are much smaller. 
In view of the computers considered to be already to small for the 
lattices used so far, it is a good idea to add to the QCD on the 
lattice the Goldstone Boson as an effective degree of freedom from 
the outside. Work in this direction is in progress \cite{Lein01}.

7) The model accounts in a trivial way for the ``non phase
transition'' behavior in the mass spectrum of mesons \cite{Isg01},
i.e. the smooth decrease of the mass splitting as a function of the
current quark mass. Due to the large masses of heavy current quarks
a meson made of heavy quarks is  not a Goldstone Boson in the strict 
sense. But, the mixing of the current-quark meson with the
constituent-quark meson in the framework of this model reproduces 
this aspect in a natural way.

8) The proposed light Goldstone $\pi$ represents an axial current.
The excited narrow nucleon states below the $\pi$ threshold will,
therefore, produce a parity violation in the $p(\vec{e},e')N^*$
reaction. Since the $N^*$ states are not resolved experimentally from
the ground state in the ongoing parity violation experiments, one has
to be careful about the interpretation of their results and the
conclusion about the strange quark content in the nucleon (for a
review see e.g. \cite{Bec01}). 

The author is indebted to Reinhard Beck, Dieter Drechsel, J\"org 
Friedrich, Stefan Scherer, Marc Vanderhaegen and Wolfram Weise for 
helpful discussions. 
This work has been supported by SFB~443 of the Deutsche 
Forschungsgemeinschaft (DFG) and the Federal State of Rhineland-Palatinate.
\def\etal{\textit{et al.}}
\def\journ#1#2#3#4{#1 \textbf{#2}, (#3) #4}
\def\EPJ#1#2#3{\journ{Eur. Phys. J.}{#1}{#2}{#3}}
\def\PRL#1#2#3{\journ{Phys. Rev. Lett.}{#1}{#2}{#3}}
\def\PR#1#2#3{\journ{Phys. Rev.}{#1}{#2}{#3}}
\def\PL#1#2#3{\journ{Phys. Lett.}{#1}{#2}{#3}}
\def\NP#1#2#3{\journ{Nucl. Phys.}{#1}{#2}{#3}}
\def\NIM#1#2#3{\journ{Nucl. Instr. and Meth.}{#1}{#2}{#3}}

\end{document}